# Generative Inverse Design of Cold Metals for Low-Power Electronics

Kedeng Wu[a], Yucheng Zhu[a], Yan Chen[b], Bizhu Zhang[a], Shuyu Liu[a], Xiaobin Deng[c], Yabei Wu[d,*], Liangliang Zhu[a,*], Hang Xiao[c,*]

[a] School of Chemical Engineering, Northwest University, Xi'an 710127, China

[b] Laboratory for Multiscale Mechanics and Medical Science, SV LAB, School of Aerospace, Xi'an Jiaotong University, Xi'an 710049, China

[c] Wu Jieh Yee School of Interdisciplinary Studies, Lingnan University, Tuen Mun, Hong Kong SAR, China

[d] Department of Materials Science and Engineering and Guangdong Provincial Key Lab for Computational Science and Materials Design, Southern University of Science and Technology, Shenzhen, Guangdong 518055, China

*E-mail: wuyb3@sustech.edu.cn, zhu.liangliang@nwu.edu.cn, hangxiao@ln.edu.hk.

# Abstract

Cold metals are a class of metals with an intrinsic energy gap located close to the Fermi level, which enables cold-carrier injection for steep-slope transistors and is therefore promising for low-power electronic applications. High-throughput screening has revealed 252 three-dimensional (3D) cold metals in the Materials Project database, but database searches are inherently limited to known compounds. Here we present an inverse-design workflow that generates 3D cold metals using MatterGPT, a conditional autoregressive Transformer trained on SLICES, an invertible and symmetry-invariant crystal string representation. We curate a training set of 26,309 metallic structures labeled with energy above hull and a unified band-edge distance descriptor that merges p-type and n-type cold-metal characteristics to address severe label imbalance. Property-conditioned generation targeting thermodynamic stability and 50-500 meV band-edge distances produces 148,506 unique candidates; 92.1% are successfully reconstructed to 3D structures and down-selected by symmetry, uniqueness and novelty filters, followed by high-throughput DFT validation. We identify 257 cold metals verified as novel with respect to the Materials Project database, with gaps around the Fermi level spanning 50-500 meV. First-principles phonon, electronic-structure, and work-function calculations for representative candidates confirm dynamical stability and contact-relevant work functions. Our results demonstrate that SLICES-enabled generative transformers can expand the chemical space of cold metals beyond high-throughput screening, providing a route to low-power electronic materials discovery.

# Introduction

Power dissipation has become a critical bottleneck in the continued scaling and performance improvement of complementary metal–oxide–semiconductor (CMOS) technology. In conventional metal–oxide–semiconductor field-effect transistors (MOSFETs), the thermionic emission mechanism imposes a fundamental lower bound on the subthreshold swing (SS) of ~60 mV/decade at room temperature (300 K), widely referred to as the Boltzmann tyranny [1–4]. This limit constrains the minimum operating voltage required to switch transistors on and off, making further reductions in dynamic power consumption exceedingly difficult. To circumvent this constraint, a variety of steep-slope device concepts have been proposed, including tunnel FETs, negative-capacitance FETs, and impact-ionization devices [1,5–10]. Among these approaches, the cold-source FET (CSFET) has attracted growing interest owing to its conceptual elegance: rather than modifying the channel transport mechanism, it employs a source contact material whose electronic structure inherently filters out high-energy carriers before they are injected into the channel [11,12].

The source material central to this strategy is termed a "cold metal." First proposed by Liu in 2020 [11], a cold metal is defined as a metallic material that possesses an intrinsic energy gap immediately above and/or below the Fermi level ($E_F$). Unlike a conventional metal, whose continuous density of states (DOS) around $E_F$ permits carrier occupation at all energies set by the Fermi–Dirac distribution, a cold metal features a DOS gap that suppresses the high-energy tail of the carrier distribution [11,13–15]. This energy-filtering window effectively blocks thermally excited ("hot") electrons from participating in injection, so that only carriers within a narrow, low-energy window contribute to current flow. When a cold metal is used as the source contact of a transistor, the injected carrier distribution is significantly colder than the Boltzmann distribution, enabling the device to achieve SS values well below the 60 mV/decade thermionic limit at room temperature [11].

Depending on the position of the band-edge gap relative to $E_F$, cold metals can be classified into three principal types: p-type, in which the gap lies above $E_F$ and filters high-energy electrons; n-type, in which the gap lies below $E_F$ and filters high-energy holes; and np-type, which exhibits gaps on both sides of $E_F$ and can filter both carrier types [11,16]. The original proposal by Liu demonstrated the concept using two-dimensional (2D) transition-metal dichalcogenide (TMD) metals—$NbX_2$ and $TaX_2$ (X = S, Se, Te)—whose electronic structures contain a gap above $E_F$, and quantum transport simulations showed that the resulting CSFETs could achieve promising switching efficiency with reduced power dissipation [11]. Subsequently, Zhang et al. extended the cold-metal paradigm to three-dimensional (3D) bulk materials, exploring TMD sulfides and oxides as cold-source contacts for monolayer $MoS_2$ transistors. Their simulations demonstrated that 3D cold-metal-contacted $MoS_2$ FETs can achieve an average SS of ~20 mV/decade over five decades of current and on-state currents exceeding 570 μA/μm [12].

Despite these encouraging results, early studies examined only a handful of cold-metal candidates. To systematically map the landscape of 3D cold metals, Zhang and Liu conducted a high-throughput computational screening of the Materials Project database, evaluating 133,693 entries and identifying 252 3D cold metals whose band-edge gaps fall within the technologically relevant window of 50-500 meV from the Fermi level [16]. Among these, 115 are p-type, 56 are n-type, and 81 are np-type. First-principles calculations on a subset of 30 representative materials revealed several distinctive physical characteristics. First, owing to the suppressed DOS near $E_F$, the electrical conductance of cold metals is systematically lower than that of conventional metals, consistent with their energy-filtering function. Second, depending on their specific band-edge gap characteristics, these cold metals exhibit highly tunable work functions that span an extensive range from 2.69 eV to 6.44 eV. This wide tunability provides significant leverage for optimizing energy band alignments at diverse metal–semiconductor interfaces [16]. Third, the thermoelectric properties of cold metals are markedly superior to those of ordinary metals: the Seebeck coefficients of representative materials such as ZrRuSb, $RbCu_4S_3$, CoSi, and $Sr_5Pb_3F$ are one to two orders of magnitude larger, and the thermoelectric figure of merit $zT$ reaches up to 0.6 at 1000 K for $Sr_5Pb_3F$ [16]. Moreover, device-level simulations of a monolayer $MoS_2$ CSFET with a ZrRuSb cold-metal contact demonstrated SS < 60 mV/decade over four decades of current and an on-state current exceeding 1 mA/μm at a supply voltage of 0.5 V, confirming the practical viability of 3D cold metals for low-power transistor applications [16].

Although this high-throughput screening effort established the first comprehensive dataset of 3D cold metals, it remains fundamentally constrained to materials already catalogued in existing crystallographic databases [17–20]. The accessible chemical and structural space is therefore bounded by prior experimental synthesis and computational enumeration, leaving a vast terra incognita of potential cold metals unexplored. Overcoming this limitation requires a paradigm shift from forward screening—which searches known materials for target properties—to inverse design strategies that directly generate novel crystal structures with prescribed electronic characteristics [21,22].

Recent advances in generative artificial intelligence, particularly Transformer-based large language models, have opened new avenues for property-driven materials discovery. Current sequence-based approaches primarily rely on direct Crystallographic Information File (CIF) [23] processing or text-conditioned training [24], such as the AtomGPT [25] framework, to capture structure-property relationships. These methods show promising capabilities in forward property prediction and generative design. However, they face fundamental limitations in representation and precision. Text-based descriptions and CIF-based sequences suffer from high representational redundancy and are not inherently invariant under translation, rotation, and permutation, typically requiring standardization or augmentation; otherwise, representational redundancy can hinder efficient learning of crystalline symmetries. Additionally, complex natural language prompts hinder the direct, high-precision mapping required to target specific physical properties [26,27].

To overcome the aforementioned limitations in representation methods and generation strategies, this study employs the MatterGPT [28] framework, a Transformer-based generative model that incorporates two distinguishing features. First, it utilizes the Simplified Line-Input Crystal-Encoding System (SLICES) [29] for crystal structure representation. SLICES is a string-based encoding that maintains invertibility and invariance to translational, rotational, and permutational operations, thereby eliminating the representational redundancy inherent in CIF format. Second, the model implements conditional generation by incorporating target property values encoded as continuous property embeddings and concatenated with the SLICES token embeddings. This architecture enables the autoregressive decoder to generate crystal structures conditioned on specified property constraints, facilitating targeted structure prediction for materials with desired characteristics.

We apply this framework to the inverse design of cold metals—to our knowledge, the first generative approach to this emerging materials class—moving beyond database-constrained high-throughput screening. To address the challenge of severe label imbalance from limited cold metal examples, we introduce a unified physical descriptor: the minimum band-edge distance, $E_{\Delta,min}$. This descriptor consolidates p-type and n-type cold metal characteristics into a single tractable learning objective. The model was trained on a curated dataset co-labeled with energy above hull and $E_{\Delta,min}$.

Property-conditioned generation targeting thermodynamic stability and 50-500 meV band-edge distances produces 148,506 unique SLICES candidates, of which 92.1% are successfully reconstructed into three-dimensional crystal structures. These candidates then undergo a systematic multi-step screening process—novelty assessment against the Materials Project database [30], symmetry constraint checks, structural uniqueness verification, and high-throughput density functional theory validation of energy above convex hull and electronic properties. This rigorous process yields 257 new cold metals not catalogued in the Materials Project database. These materials exhibit energy gaps around the Fermi level ranging from 50 to 500 meV and energies above the convex hull below 0.25 eV/atom, demonstrating near-hull stability and potential synthesizability. First-principles phonon dispersion calculations confirm dynamical stability of representative candidates, while electronic structure and work function analysis validates their cold metal characteristics and suitability for semiconductor contacts. These results demonstrate that SLICES-enabled generative Transformers can effectively expand the accessible chemical space of cold metals beyond high-throughput screening, providing a computational route to discover low-power electronic materials.

# Methods

## SLICES representation

To enable effective generative modeling of crystalline materials, we employ SLICES, a string-based representation scheme that addresses the fundamental challenges of encoding infinite periodic structures. Similar to how SMILES [31] notation revolutionized molecular informatics by converting molecular graphs into linear strings, SLICES transforms three-dimensional crystal structures into compact, unambiguous text sequences. SLICES possesses two defining characteristics: invertibility, which enables accurate reconstruction of the original structure, and invariance, ensuring that the representation remains unchanged under symmetry operations including translation, rotation, and atom permutation. These characteristics make SLICES particularly well-suited for autoregressive generative models like Transformers [32,33], which excel at learning patterns in sequential data.

The encoding workflow transforms crystalline structures through labeled quotient graphs, finite representations that capture the essential topology and periodicity of infinite lattices. As illustrated in **Fig. 1**, this conversion proceeds in three stages: (1) crystal structure files are parsed into structure objects using Pymatgen [34]. (2) bonding networks are established via the CrystalNN algorithm [35], which identifies atomic connectivity based on coordination environments. (3) the resulting connectivity graph is converted into a SLICES string by encoding elemental composition, node indices for atoms within the unit cell, and critically, edge labels that specify translational periodicity. Each edge is represented as "i j m n p", where indices i and j identify connected atoms, while the triplet m n p defines the lattice translation vector linking unit cells. To prevent ambiguity during parsing, characters 'o', '+', and '−' denote 0, +1, and −1 in edge labels, ensuring numerical digits exclusively indicate node indices. Symmetry information is incorporated by appending a modified Pymatgen "enc"-style space group string to the SLICES representation, where numeric characters are replaced by letters to avoid conflicts with node indices. This encoding scheme elegantly captures both chemical composition and crystallographic periodicity in a format accessible to sequence-based models.

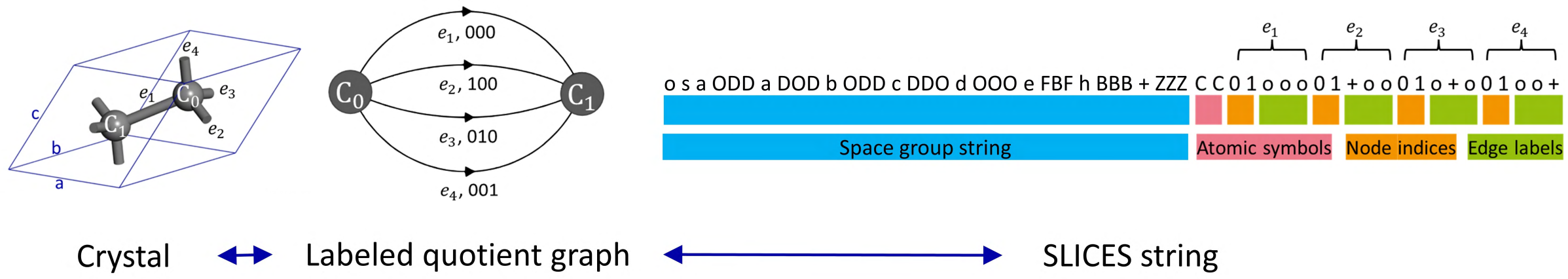

**Fig. 1** | Conversion from a 3D crystal structure to a SLICES string

Reconstruction from SLICES strings to three-dimensional structures is accomplished through the SLI2Cry algorithm, a multi-stage refinement process. Initial structures are generated using Eon's topology-based method, which leverages graph theory and the encoded edge labels to establish correct connectivity and periodicity. These preliminary geometries then undergo optimization based on chemically meaningful geometry predicted with a modified GFN-FF [36] force field, which estimates equilibrium bond lengths and angles. Finally, the M3GNet [37] universal interatomic potential performs structural refinement. Benchmarking on diverse crystal datasets demonstrates that this three-stage approach achieves 94.95% reconstruction success, validating SLICES as a robust foundation for generative materials design.

# MatterGPT architecture

For inverse design of cold metal materials, we employ MatterGPT [28], an autoregressive Transformer architecture operating on SLICES representations. **Fig. 2** illustrates the workflow: during training, the model learns from crystal structures paired with DFT-calculated properties to establish the conditional probability distribution P(SLICES|$p_1$, ..., $p_n$); during generation, this distribution guides creation of novel SLICES strings conditioned on target properties. Generated strings are reconstructed into crystal structures via SLI2Cry and subjected to computational screening.

MatterGPT adopts a decoder-only architecture with 12 stacked Transformer blocks (**Fig. 2b**). Input SLICES strings undergo preprocessing where a start token '>' is prepended and padding tokens '<' appended to reach uniform length N. The SLICES vocabulary comprises 132 tokens: 83 atomic symbols, 20 node indices, 27 edge labels, and 2 special tokens. Property specifications ($p_1$, ..., $p_n$) are concatenated with preprocessed SLICES sequences to form the input.

The input representation integrates three embedding components through element-wise summation: token embeddings encoding vocabulary elements, positional embeddings preserving sequential order, and type embeddings distinguishing properties (type = 0) from SLICES tokens (type = 1). Each block contains a masked multi-head self-attention module followed by a feed-forward network, with residual connections and layer normalization applied around each sub-layer. The core self-attention mechanism implements scaled dot-product attention:

$$Attention(Q, K, V) = softmax\left(\frac{QK^T}{\sqrt{d_k}}\right)V \quad (1)$$

where Q, K, and V represent query, key, and value matrices derived from input embeddings through learned linear transformations, and d denotes the dimension of the key vectors. The masking mechanism restricts each token to attend only to preceding positions, ensuring autoregressive generation. The architecture employs 12 attention heads operating in parallel, each with 64-dimensional projections. Outputs from all heads are concatenated and linearly transformed, enabling the model to jointly capture information from different representation subspaces. The complete architecture comprises 768-dimensional embeddings and 3072-dimensional feed-forward networks, approximately 80 million trainable parameters.

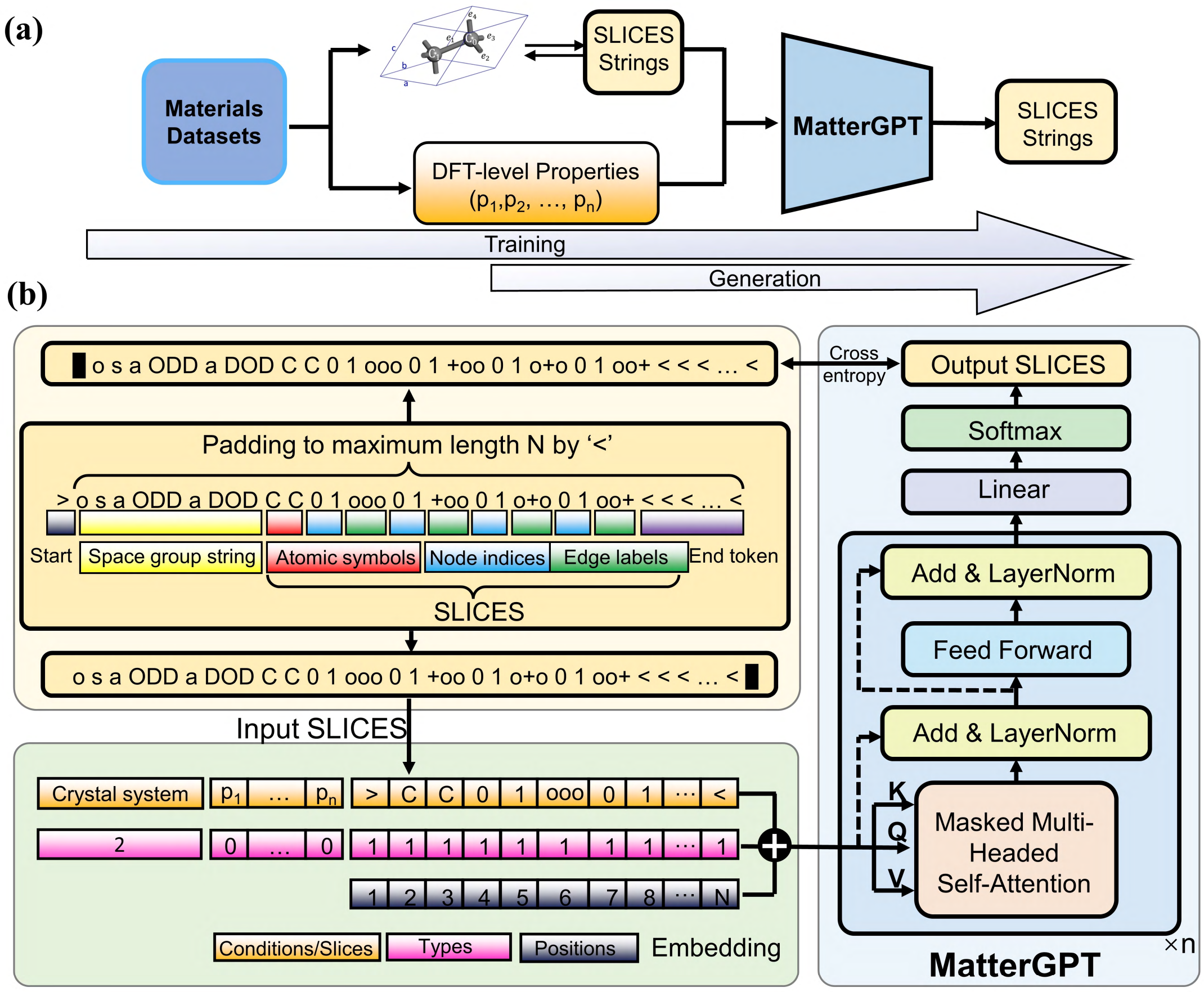

**Fig. 2** | The architecture and workflow of MatterGPT. **(a)** Training and generative sampling pipeline of MatterGPT. **(b)** Detailed model architecture and SLICES string preprocessing workflow. Crystal system, property, and SLICES token embeddings are each assigned type embeddings (0 = properties, 1 = SLICES tokens, 2 = crystal system) to distinguish input modalities, and combined with trainable positional embeddings to capture sequential dependencies.

Training proceeds by maximizing the conditional probability of SLICES sequences given target properties through cross-entropy loss minimization, using the Adam optimizer with a learning rate of 0.0001 and batch size of 60. During generation, Gumbel-Softmax sampling with temperature 1.2 introduces controlled stochasticity to the predicted token distributions. We complement this with nucleus sampling at a cumulative probability threshold of 0.9, restricting token selection to the top 90% probability mass to maintain both generation diversity and property-directed coherence.

## First-principles calculation details

All density functional theory (DFT) [38,39] calculations were performed using the Vienna Ab initio Simulation Package (VASP) [40], employing the projector augmented wave (PAW) method [41] to describe core-valence interactions. The exchange-correlation functional was treated within the generalized gradient approximation (GGA) using the Perdew-Burke-Ernzerhof (PBE) parameterization [42]. To ensure the convergence of total energy, a plane-wave basis set with a kinetic energy cutoff of 520 eV was uniformly applied.

Structural geometry optimization was conducted through a rigorous three-stage relaxation protocol: initially adjusting ionic positions and cell volume, followed by cell shape optimization, and concluding with the refinement of ionic coordinates. This process continued until the residual forces on all atoms converged to less than 0.01 eV/Å. For Brillouin zone integration, we employed Monkhorst-Pack k-point grids generated with a reciprocal space density of 32 to guarantee both computational accuracy and efficiency.

To evaluate the dynamical stability of the identified cold metals, phonon dispersion spectra were computed using the finite displacement method as implemented in the Phonopy package [43]. For these calculations, the kinetic energy cutoff was set to 500 eV, and a stringent energy convergence criterion of $10^{-8}$ eV was applied to the electronic self-consistency loop to ensure high precision in the resulting force constants.

Furthermore, to assess the electronic energy level alignment critical for device applications, we calculated the work functions ($\Phi$) of representative candidates [44]. Slab models were constructed with a vacuum layer exceeding 15 Å along the surface normal to eliminate spurious interactions between periodic images. The work function was determined via the relation:

$$\Phi = E_{vac} - E_F \tag{2}$$

where $E_F$ denotes the Fermi level obtained from the electronic structure calculation, and $E_{vac}$ represents the vacuum energy level obtained from the plateau of the planar-averaged electrostatic potential in the vacuum region.

# Results

## Cold metal screening

Our inverse design workflow begins with the assembly of a comprehensive training dataset from the Materials Project (MP) database [30]. We focused specifically on cold metals, a distinct class of materials characterized by an intrinsic energy gap near the Fermi level. This electronic feature facilitates the effective filtering of high-energy carriers, thereby offering a promising pathway to minimize power dissipation in low-power electronic devices.

Following the classification framework established by Zhang et al. [16], we categorize cold metals into three principal types based on the position of their band-edge gaps relative to the Fermi level, as schematically illustrated by the density of states (DOS) diagrams in **Fig.** 3. p-type cold metals (**Fig. 3a**), analogous to heavily p-doped semiconductors, are characterized by an energy gap positioned above the Fermi level. The key descriptor for this class is $E_{CD}$, defined as the energy difference between the conduction band edge of this gap and the Fermi level. n-type cold metals (**Fig. 3b**), which resemble heavily n-doped semiconductors, feature an energy gap located below the Fermi level. Their defining parameter is $E_{VD}$, representing the energy difference between the Fermi level and the valence band edge of this gap. np-type cold metals (**Fig. 3c**) exhibit the presence of gaps both above and below the Fermi level and are thus characterized by both an $E_{CD}$ and an $E_{VD}$.

Building on the observation that cold metal properties converge toward conventional metallic behavior as the gap-edge distance increases, we adopt the screening criterion established by Zhang [16]: materials are classified as cold metals when the energy difference between the band-edge gap and the Fermi level falls within the range of 50 meV to 500 meV. This energy window captures the regime where carrier filtering effects are most pronounced while maintaining sufficient metallic conductivity for device applications.

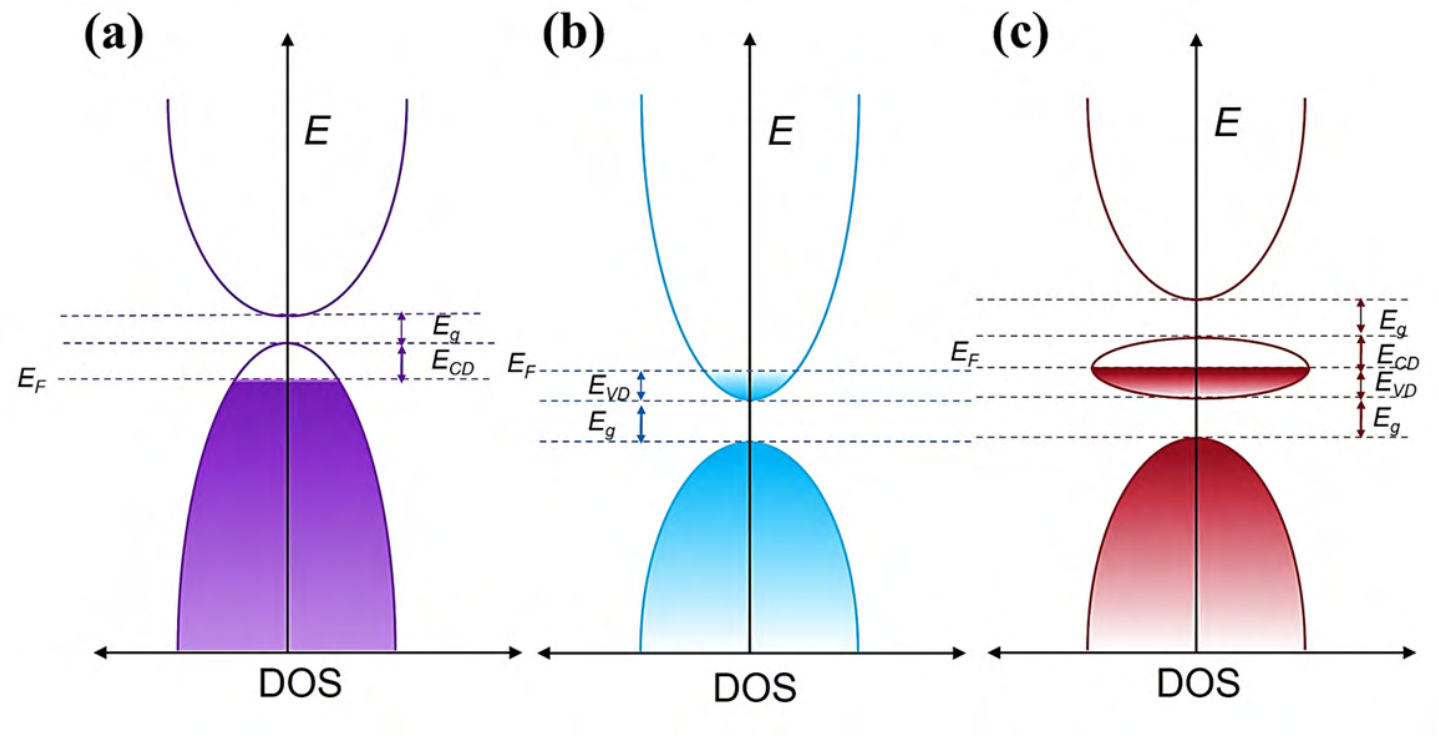

**Fig. 3 |** Schematic density of states diagrams illustrating the three principal types of cold metals. (**a**) p-type cold metal, analogous to a heavily p-doped semiconductor, characterized by an energy gap positioned above the $E_F$. (**b**) n-type cold metal, analogous to a heavily n-doped semiconductor, exhibiting an energy gap located below $E_F$. (**c**) np-type cold metal, featuring energy gaps both above and below $E_F$, characterized by both $E_{CD}$ and $E_{VD}$.

To construct the training dataset, we extracted metallic materials from the Materials Project database and applied a rigorous four-step screening protocol:

(1) Only three-dimensional structures were retained, as the labeled quotient graph methodology integral to SLICES encoding is applicable exclusively to 3D periodic crystals.

(2) The dataset was constrained to materials composed of elements with atomic numbers not exceeding 86, accounting for the operational limits of the modified GFN-FF [36] force field in the SLICES reconstruction algorithm. This threshold ensures both stability and accuracy during structural reconstruction.

(3) All non-metallic structures were excluded to focus the analysis exclusively on metallic materials, consistent with the definition of cold metals as a specialized metallic class.

(4) Structures containing more than 20 atoms per unit cell were removed, following established practice in SLICES-based generative modeling frameworks to maintain computational efficiency and training dataset consistency.

These filtering criteria resulted in a final dataset of 26,309 metallic materials. To avoid the data scarcity associated with restricting the study to known cold metals, we included all metallic structures and calculated their $E_{CD}$ and $E_{VD}$ values. This expansion enables the model to capture features from a wider chemical space. However, the direct application of these properties as training targets exposed a fundamental challenge arising from severe distributional imbalances. As illustrated in **Fig. 4**, the raw distributions of $E_{CD}$ and $E_{VD}$ reveal two distinct patterns that present significant challenges for model training. The $E_{CD}$ distribution is heavily skewed toward zero. Since most metallic materials lack an energy gap above the Fermi level, over 21,000 samples exhibit an $E_{CD}$ of 0. In contrast, samples with non-zero $E_{CD}$ values are sparsely distributed across the range greater than 0 eV, resulting in limited data density for learning meaningful features in these regions. Conversely, the $E_{VD}$ distribution shows an opposing imbalance. While most metals have electronic states below the Fermi level, their $E_{VD}$ values are predominantly concentrated in the high-energy region (>3 eV), which accounts for over 22,000 samples. This distribution is problematic because the region relevant for cold metal applications and the intermediate range (0–3 eV) are severely underrepresented, containing only a small fraction of the dataset.

To address these distributional imbalances, we introduce a unified descriptor termed the minimum band-edge distance, $E_{\Delta,min}$, defined as:

$$E_{\Delta,min} = \begin{cases} min(|E_{CD}|, |E_{VD}|), & E_{CD} \neq 0 \; and \; E_{VD} \neq 0 \\ |E_{CD}|, & E_{VD} = 0 \\ |E_{VD}|, & E_{CD} = 0 \end{cases} \quad (3)$$

where $E_{CD}$ and $E_{VD}$ denote the energy distances from $E_F$ to the nearest edge of the cold-metal gap on the conduction side ($E>E_F$) and valence side ($E<E_F$), respectively. In this formulation, the minimization is conditional: if a gap is absent on one side (i.e., the value is 0), that component is excluded to isolate the active energy-filtering window. Consequently, for single-sided cold metals, $E_{\Delta,min}$ reduces to the non-zero gap value. This construction naturally focuses training on materials with band-edge features near the Fermi level. As shown by the magenta bars in **Fig. 4**, $E_{\Delta,min}$ exhibits a more balanced distribution. Compared to individual $E_{CD}$ and $E_{VD}$ distributions, $E_{\Delta,min}$ provides substantially improved coverage in the low-energy regime (0–3 eV), with especially enhanced representation in the 0–0.5 eV range that contains the cold-metal screening window (50-500 meV).

We evaluated generative performance by training MatterGPT models on three property combinations: energy above hull (EAH) - $E_{CD}$, EAH - $E_{VD}$, and the unified EAH-$E_{\Delta,min}$, generating equal numbers of structures from each. As shown in Table 1, $E_{\Delta,min}$ exhibits the highest uniqueness (84.40%) and novelty (41.04%) among the three descriptors. Although it displays marginally lower symmetry (35.04%) than $E_{VD}$ (36.19%), its superior performance in uniqueness and novelty justifies this minor trade-off. Based on these results, we constructed the final training dataset of 26,309 metallic materials labeled with EAH and $E_{\Delta,min}$. Training MatterGPT on this dataset enables systematic generation of cold metals with targeted electronic properties biased toward thermodynamic stability.

**Table 1.** Comparison of properties under different descriptors

| Descriptor | Performance (%) | | |
|---|---|---|---|
| | Uniqueness | Symmetry | Novelty |
| $E_{CD}$ | 79.30 | 16.36 | 32.87 |
| $E_{VD}$ | 77.84 | 36.19 | 32.16 |
| $E_{\Delta,min}$ | **84.40** | 35.04 | **41.04** |

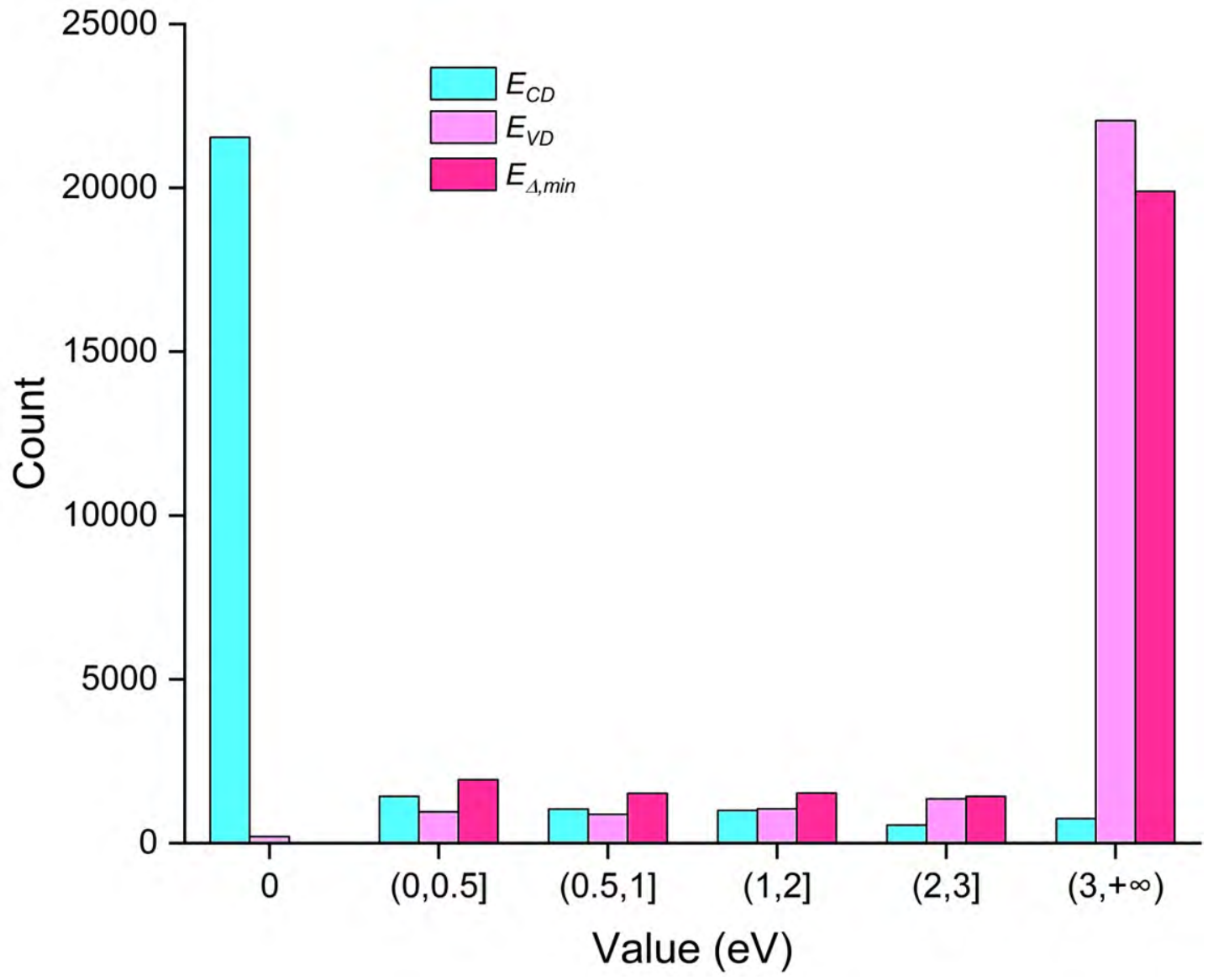

**Fig. 4 |** $E_{CD}$ , $E_{VD}$ and $E_{\Delta,min}$ distribution in the dataset. The null count for $E_{\Delta,min}$ at 0 eV reflects the absence of materials in the dataset where both $E_{CD}$ and $E_{VD}$ vanish simultaneously.

## Discovery of new cold metal materials

Utilizing the trained MatterGPT model, we performed inverse design to generate cold metal candidates, targeting thermodynamic stability and $E_{\Delta,min}$ within the 50-500 meV range. As illustrated in **Fig. 5**, the workflow begins by embedding these target properties, which condition the model to autoregressively generate SLICES sequences token by token. This process yielded 148,506 unique SLICES strings, each encoding the compositional and topological features of a potential cold metal. Subsequently, we converted these linear sequences into three-dimensional crystal structures using the SLI2Cry reconstruction algorithm. This conversion achieved a high success rate of 92.1%, resulting in 136,766 valid structures. The few reconstruction failures were primarily attributed to invalid edge labels or topological inconsistencies in the generated sequences.

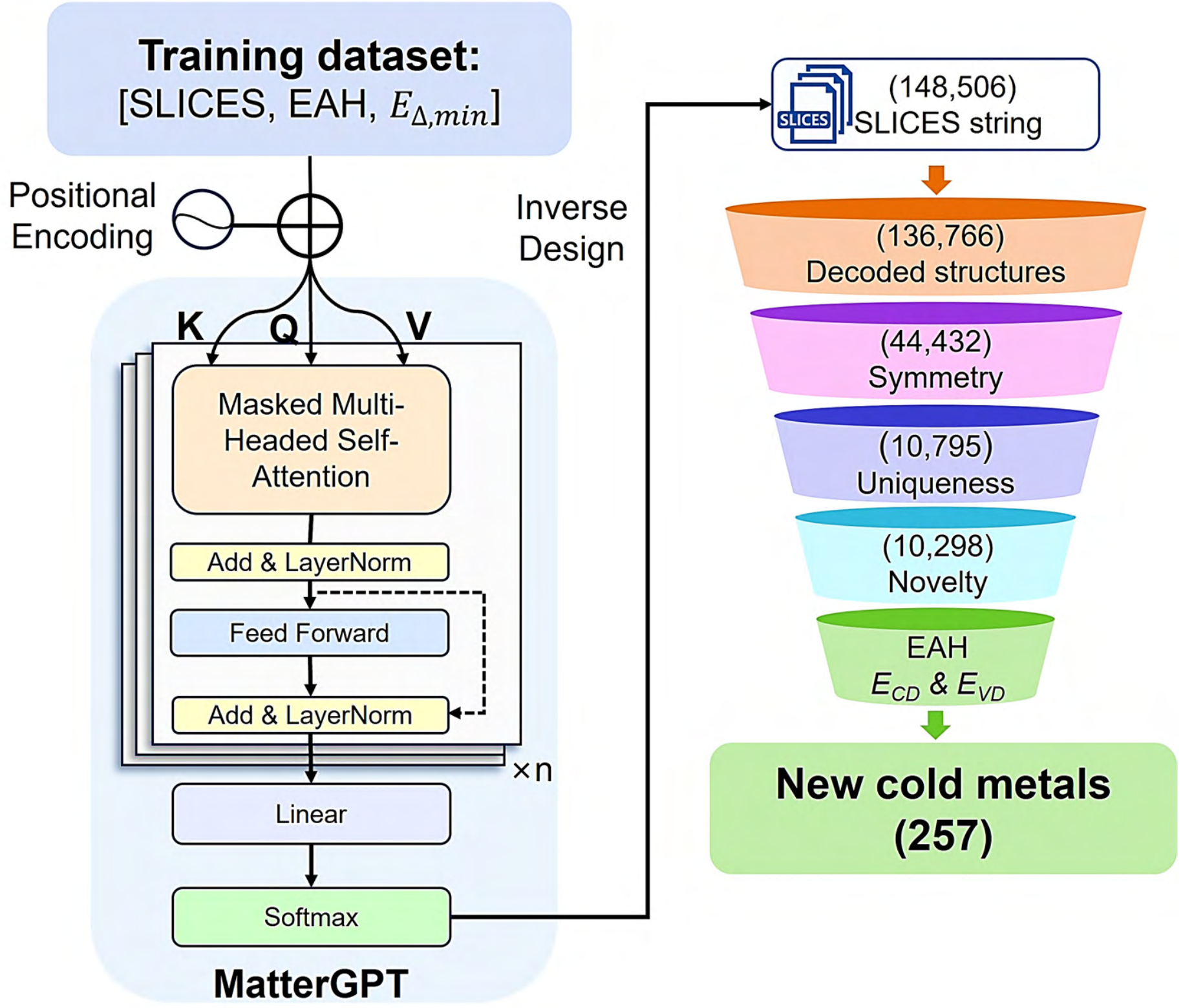

**Fig. 5 |** Schematic overview of the MatterGPT-based inverse design workflow for generating novel cold metal candidates. Target properties, including EAH and $E_{\Delta,min}$, are embedded as conditioning inputs to guide the autoregressive generation of SLICES sequences. These sequences are subsequently converted into three-dimensional crystal structures via the reconstruction algorithm. The resulting candidates are then subjected to a three-stage screening protocol encompassing symmetry filtering, uniqueness verification, and novelty assessment against the Materials Project database, ultimately affording a pool of structurally valid and chemically novel cold metal candidates for further evaluation.

The reconstructed candidates were subjected to a rigorous three-stage screening protocol to ensure structural validity and novelty. First, we excluded structures assigned to the P1 space group to prioritize candidates with relatively high crystallographic symmetry for downstream analysis. Second, we evaluated uniqueness to eliminate redundant candidates produced within the same generative run. Following the methodology of Zeni et al. [45], we compared structures based on space groups and composition-normalized atomic coordinates—a robust approach dealing with both ordered and disordered configurations. Finally, we assessed novelty by screening against the Materials Project [30] database. This verification ensures that the generated candidates represent previously unreported materials rather than reproductions of entries already catalogued in this comprehensive repository. This multi-stage pipeline refined the initial pool to 10,298 entirely novel candidates. All computational tasks, from training to generation, were performed on a workstation equipped with an Intel i9-10900KF processor and an NVIDIA RTX 3090 GPU (24 GB VRAM).

To evaluate the energy above hull and intrinsic cold metal characteristics of the 10,298 novel candidate structures, we performed high-throughput DFT calculations using VASP. This comprehensive assessment involved computing the energy above hull relative to known phases in the Materials Project database to quantify thermodynamic stability, while concurrently determining $E_{CD}$ and $E_{VD}$ values to confirm definitive cold metal features. Following Mal et al. [46], we retained only materials with EAH < 0.25 eV/atom, a threshold that identifies metastable yet experimentally accessible compounds. This final validation step reduced the candidate pool to 257 cold metals.

All validated candidates exhibit $E_{CD}$ and/or $E_{VD}$ values within the targeted 50-500 meV window, supporting their assignment as p-type, n-type, or np-type cold metals. Tables 2–4 present representative examples, whereas the complete validated data for these candidates are provided in Tables S1–S3 of the Supplementary Information. Notably, our generative approach differs fundamentally from previous high-throughput screening methods. While Zhang et al. [16] identified 252 cold metals by screening the Materials Project database, their strategy was inherently confined to previously documented materials. In contrast, our MatterGPT-based framework yielded 257 entirely novel compounds.

**Table 2.** p-type cold metals among candidate materials

| Formula | EAH (eV/atom) | $E_{CD}$ (eV) | $E_{VD}$ (eV) | Crystal system |
|---|---|---|---|---|
| $KTm_2Sb2Se_7$ | 0.094 | 0.208 | 4.645 | Orthorhombic |
| $Na_2CuRuF_6$ | 0.197 | 0.076 | 1.320 | Cubic |
| $CsBaF_4$ | 0.078 | 0.141 | 1.734 | Tetragonal |
| $CsNa_2ZrI_8$ | 0.097 | 0.267 | 3.371 | Trigonal |
| $Sr_2EuCoO_6$ | 0.142 | 0.358 | 2.400 | Cubic |

**Table 3.** n-type cold metals among candidate materials

| Formula | EAH (eV/atom) | $E_{CD}$ (eV) | $E_{VD}$ (eV) | Crystal system |
|---|---|---|---|---|
| $Nb_2N_3OF_{12}$ | 0.081 | 0.728 | 0.334 | Monoclinic |
| $Rb_6SrCeCl_{12}$ | 0.104 | 2.258 | 0.116 | Monoclinic |
| $Na_3MnCl_5$ | 0.107 | 2.614 | 0.278 | Cubic |
| $Cs_2MnH_7$ | 0.052 | 1.201 | 0.300 | Cubic |
| $Rb_3SeBr_5$ | 0.163 | 1.127 | 0.202 | Tetragonal |

**Table 4.** np-type cold metals among candidate materials

| Formula | EAH (eV/atom) | $E_{CD}$ (eV) | $E_{VD}$ (eV) | Crystal system |
|---|---|---|---|---|
| $Cs_2TiCoF_6$ | 0.068 | 0.143 | 0.083 | Cubic |
| $Cs_5CeF_{14}$ | 0.038 | 0.075 | 0.090 | Monoclinic |
| $YbCuF_6$ | 0.141 | 0.132 | 0.191 | Trigonal |
| $Na_2TiNbCl_8$ | 0.091 | 0.240 | 0.349 | Monoclinic |
| $K_2HgBiF_6$ | 0.226 | 0.441 | 0.123 | Cubic |

# Dynamic stability verification and electronic structure analysis

To validate the predicted properties and assess device application potential, we selected two representative p-type cold metal candidates from the 257 compounds discovered: $CsBaF_4$ ($E_{CD}$ = 0.14 eV, EAH = 0.08 eV/atom) and $RbBaSe_2$ ($E_{CD}$ = 0.27 eV, EAH = 0.22 eV/atom). We performed phonon dispersion calculations using the finite displacement method implemented in Phonopy [43] to confirm dynamical stability, following the computational protocols described in Section 2.3. Electronic band structure calculations were conducted to verify their intrinsic cold metal characteristics.

To further assess the suitability of the selected candidates for contact engineering, we evaluated their work functions using low-index surface slab models with symmetry-preserving terminations, since interfacial energy-level alignment is a key factor governing metal–semiconductor contact behavior [47]. In all calculations, vacuum regions thicker than 15 Å were introduced to suppress spurious interactions between periodic images. The work function was determined from the difference between the vacuum level, extracted from the planar-averaged electrostatic potential in the vacuum region, and the Fermi level. As summarized in **Fig. 6**, both compounds exhibit contact-relevant work functions together with robust dynamical stability and the characteristic electronic signature of p-type cold metals. Specifically, the phonon dispersion curves in **Fig. 6c** and **Fig. 6g** show no imaginary modes throughout the Brillouin zone [48], confirming dynamical stability at 0 K, while the corresponding band structures in **Fig. 6b** and **Fig. 6f** reveal a distinct gap in the unoccupied states above the Fermi level, consistent with the defining feature of p-type cold metals.

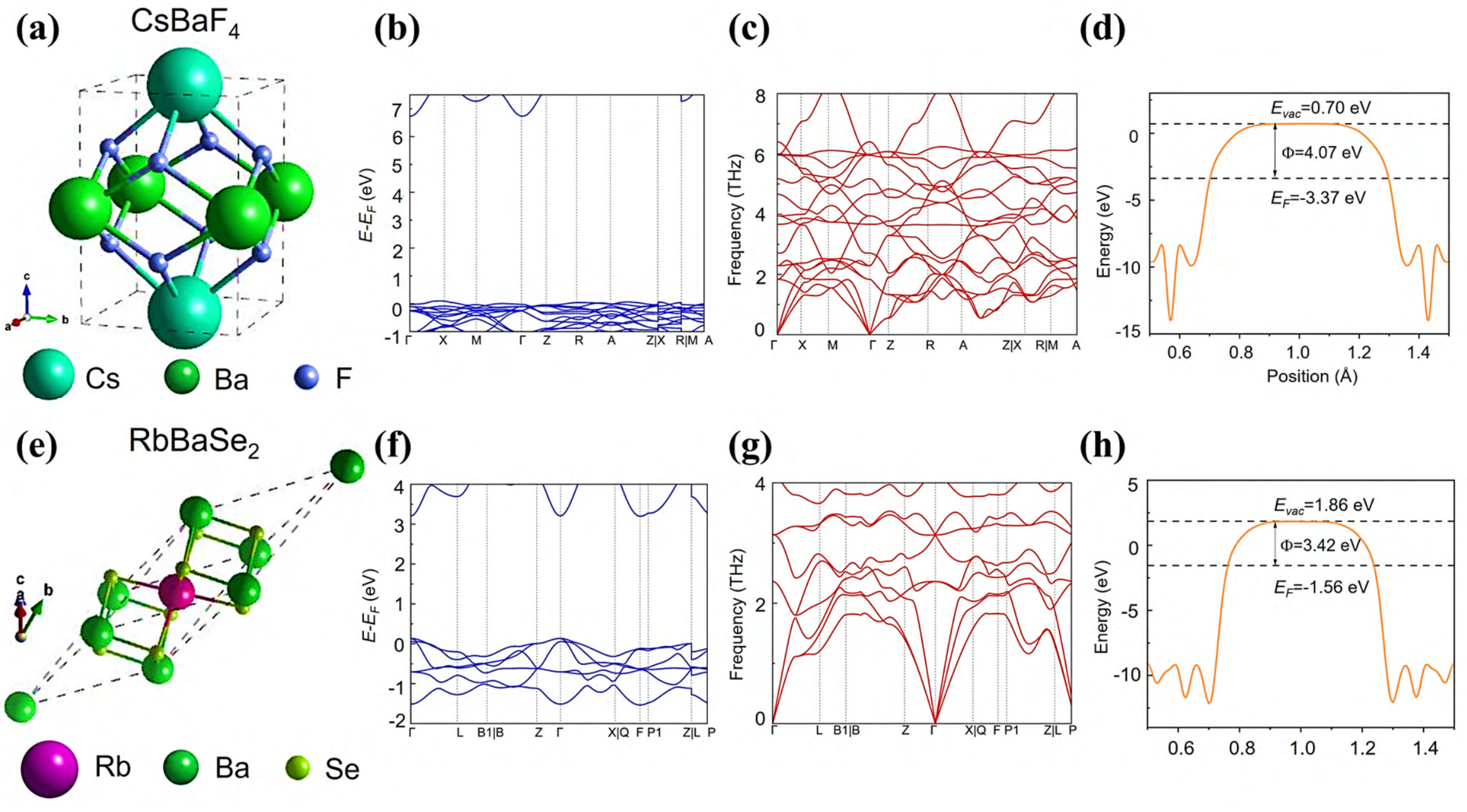

**Fig. 6** | Structural, dynamical, and electronic properties of two representative p-type cold metals. **(a)** Crystal structure of $CsBaF_4$. **(b)** Electronic band structure of $CsBaF_4$. **(c)** Phonon dispersion of $CsBaF_4$. **(d)** Planar-averaged electrostatic potential of a representative surface slab of $CsBaF_4$, yielding a work function of 4.07 eV. **(e)** Crystal structure of $RbBaSe_2$. **(f)** Electronic band structure of $RbBaSe_2$. **(g)** Phonon dispersion of $RbBaSe_2$. **(h)** Planar-averaged electrostatic potential of a representative surface slab of $RbBaSe_2$, yielding a work function of 3.41 eV. The dashed lines indicate the vacuum level and Fermi level.

Work function analysis yields $\Phi$ = 4.07 eV for $CsBaF_4$ (**Fig. 6d**) and $\Phi$ = 3.41 eV for $RbBaSe_2$ (**Fig. 6h**), respectively. These values fall comfortably within the work function window for cold-electron injection, aligning perfectly with the established energetic regimes of cold metals reported in recent literature [16]. Notably, they closely match that of ZrRuSb (3.6 eV), which achieves subthreshold switching of 60 mV/decade in $MoS_2$ transistors [16]. This suggests similar potential for $CsBaF_4$ and $RbBaSe_2$ in cold electron injection applications. Their intrinsic band gaps of 50-500 meV above the Fermi level effectively filter high-energy carriers. This combination of favorable work functions and band-edge gap positioning makes $CsBaF_4$ and $RbBaSe_2$ strong candidates for steep-slope transistors and other low-power electronic devices. These findings validate that MatterGPT can generate novel cold metals with electronic properties matching those from high-throughput screening.

# Conclusion

We present an automated workflow for the inverse design of three-dimensional cold metals through generative artificial intelligence. In contrast to conventional high-throughput screening methods limited by finite material databases, our generative framework directly generates novel crystal structures with targeted electronic properties. We addressed the challenge of sparse training data by introducing a unified physical descriptor, the minimum band-edge distance $E_{\Delta,min}$. This descriptor consolidates the identification of p-type, n-type, and np-type cold metals into a single learning objective, enabling the MatterGPT model to generate target materials more efficiently.

By targeting thermodynamic stability and specific $E_{\Delta,min}$ values, the model generated 148,506 candidate SLICES strings. We applied a multi-stage screening protocol involving structural reconstruction, novelty assessment, symmetry filtering, and high-throughput DFT calculations. This approach yielded 257 cold metal candidates with near-hull stability, none of which were matched to entries in the Materials Project database at the time of screening. First-principles validation of two representative p-type materials, $CsBaF_4$ and $RbBaSe_2$, confirmed their dynamical stability via phonon dispersion analysis and verified their characteristic cold metal electronic structures. Both materials exhibit work functions of 4.07 eV and 3.41 eV, respectively, within the optimal range for semiconductor contact applications.

These results confirm that our approach can discover functional materials with targeted properties. The work makes two important contributions. It substantially expands the library of candidate cold metals for electronic applications, while their thermoelectric potential warrants future dedicated investigation. It also provides a viable computational framework for inverse design in sparsely populated target-property spaces. Future work should prioritize synthesis and interface characterization of the most stable candidates, while improving generative diversity and accuracy through pre-training on larger crystal datasets and iterative DFT-in-the-loop refinement.

# Declaration of Interest Statement

The authors declare that they have no known competing financial interests or personal relationships that could have appeared to influence the work reported in this paper.

# Acknowledgments

This work was financially supported by the National Key Research and Development Program of China (Grant No.: 2024YFA1209801), the National Natural Science Foundation of China (Grant Nos.: 22203066, 12302140, and 12574265), the China Postdoctoral Science Foundation (Grant Nos.: 2023M732794 and 2025T180517), the Fundamental Research Funds for the Central Universities (Grant No.: sxzy012023213), and the Special Fund for AI-Empowered Scientific Research of the Shaanxi Provincial Department of Science and Technology (Grant No.: 2025YXYC012).

# Data availability

All data needed to evaluate the conclusions in the paper are present in the paper and/or the Supplementary Materials. The scripts that can generate the datasets, weights of the trained model, and related code can be found in the permanent Zenodo repository (https://doi.org/10.5281/zenodo.18981860).

# Code availability

The SLICES source code is available on GitHub (https://github.com/xiaohang007/SLICES). SLICES v2.0 was used in this work. Code and data for MatterGPT are available on GitHub (https://github.com/xiaohang007/SLICES/tree/main/MatterGPT/).

# Supplementary information

**Table S1. Complete list of validated p-type cold metal candidates**

| Number | Formula | EAH (eV/atom) | $E_{CD}$ (eV) | $E_{VD}$ (eV) | $E_F$ (eV) | Crystal system |
|---|---|---|---|---|---|---|
| 1 | $RbYbSe_2$ | 0.1912 | 0.3120 | 2.6389 | 0.4548 | trigonal |
| 2 | $Ba_2YCoO_6$ | 0.1250 | 0.2764 | 2.1062 | 2.4463 | cubic |
| 3 | $NaYbSe_2$ | 0.1705 | 0.3971 | 3.1675 | 0.9318 | tetragonal |
| 4 | $Cs_2RbYbCl_6$ | 0.1374 | 0.0815 | 0.6326 | -1.8893 | cubic |
| 5 | $Ba_2PrCoO_6$ | 0.1561 | 0.3387 | 2.0739 | 2.4639 | cubic |
| 6 | $Cs_3YbCl_6$ | 0.1290 | 0.0863 | 1.4304 | -1.8717 | cubic |
| 7 | $Ba_2BiBr_2$ | 0.2209 | 0.3822 | 2.0353 | 2.4615 | tetragonal |
| 8 | $Ba_2NaVO_6$ | 0.2034 | 0.3724 | 3.9167 | 1.5095 | cubic |
| 9 | $SrNdMnReO_6$ | 0.1806 | 0.3543 | 1.2130 | 5.5264 | cubic |
| 10 | $Na_3CeSe_6$ | 0.1787 | 0.2420 | 2.2157 | 1.7431 | cubic |
| 11 | $K_2RuBr_7$ | 0.1354 | 0.2016 | 2.4339 | -0.1408 | cubic |
| 12 | $K_2EuTlCl_6$ | 0.0522 | 0.4394 | 0.7296 | -0.4977 | cubic |
| 13 | $Na_3NbH_6$ | 0.1900 | 0.3104 | 0.5781 | 0.6140 | cubic |
| 14 | $Rb_2NaCCl_6$ | 0.1118 | 0.1445 | 2.2667 | -1.4935 | cubic |
| 15 | $Cs_2KSrH_6$ | 0.2455 | 0.1567 | 2.3722 | -1.1075 | cubic |
| 16 | $Ba_2EuCoO_6$ | 0.0949 | 0.2662 | 2.2637 | 2.4348 | cubic |
| 17 | $Na_2PdF_7$ | 0.2290 | 0.1026 | 0.5909 | -2.2464 | cubic |
| 18 | $K_2RbCaCl_6$ | 0.1518 | 0.0861 | 1.9245 | -2.4118 | cubic |
| 19 | $KBa_2BiN$ | 0.2339 | 0.3716 | 2.0601 | 2.2023 | cubic |
| 20 | $K_2TlFeCl_6$ | 0.1385 | 0.4281 | 1.0549 | 1.3695 | cubic |
| 21 | $Cs_2NaSrCl_6$ | 0.1121 | 0.1192 | 1.9573 | -1.4327 | cubic |
| 22 | $NbMo_6Se_8$ | 0.1647 | 0.2351 | 6.3926 | 5.8664 | trigonal |
| 23 | $Rb_2TlFeBr_6$ | 0.1243 | 0.4341 | 0.9389 | 0.9000 | cubic |
| 24 | $YbScF_6$ | 0.2371 | 0.1119 | 2.5083 | -4.7933 | trigonal |
| 25 | $K_2PdBr_7$ | 0.0753 | 0.2461 | 2.2820 | -0.1768 | cubic |
| 26 | $Sr_2CaVO_6$ | 0.1608 | 0.1477 | 1.7775 | 1.8997 | cubic |

| Number | Formula | EAH (eV/atom) | $E_{CD}$ (eV) | $E_{VD}$ (eV) | $E_F$ (eV) | Crystal system |
|---|---|---|---|---|---|---|
| 27 | $Cs_2KYbCl_6$ | 0.1161 | 0.0972 | 0.7052 | -1.7380 | cubic |
| 28 | $Sr_2TbMnO_6$ | 0.0890 | 0.0923 | 2.1133 | 2.0609 | cubic |
| 29 | $TaInIr_1$ | 0.2332 | 0.3214 | 4.9135 | 6.6982 | cubic |
| 30 | $KSrSe_2$ | 0.2182 | 0.2662 | 2.0410 | 0.1556 | trigonal |
| 31 | $Cs_2YbTlF_6$ | 0.0763 | 0.4503 | 0.6295 | 0.9554 | cubic |
| 32 | $BaZrV_2O_6$ | 0.1384 | 0.3209 | 1.0076 | 5.0887 | cubic |
| 33 | $Ba_2SmCoO_6$ | 0.1408 | 0.2941 | 2.0663 | 2.3440 | cubic |
| 34 | $K_2SrTlF_6$ | 0.1705 | 0.4337 | 0.5368 | 0.4672 | cubic |
| 35 | $K_2RuI_7$ | 0.1720 | 0.1507 | 2.3409 | 1.6080 | cubic |
| 36 | $NaYbSe_2$ | 0.1529 | 0.2600 | 3.0033 | 0.8659 | trigonal |
| 37 | $Sr_2YbVO_6$ | 0.1761 | 0.1446 | 4.1439 | 1.6855 | cubic |
| 38 | $Sr_3CoO_6$ | 0.2404 | 0.3819 | 2.0749 | 1.9390 | cubic |
| 39 | $Sr_3NiO_6$ | 0.1664 | 0.4818 | 2.0290 | 1.8406 | cubic |
| 40 | $Rb_2VCl_7$ | 0.1432 | 0.2086 | 0.5295 | -0.0268 | cubic |
| 41 | $K_2EuTlCl_6$ | 0.0316 | 0.0852 | 2.1590 | -1.7862 | cubic |
| 42 | $Ba_2SrCoO_6$ | 0.1745 | 0.2945 | 1.9314 | 2.0786 | cubic |
| 43 | $Cs_2RbCaCl_6$ | 0.1374 | 0.0796 | 0.6400 | -1.8207 | cubic |
| 44 | $Ba_2NdCoO_6$ | 0.1476 | 0.2937 | 2.0494 | 2.3591 | cubic |
| 45 | $K_2TiCuCl_6$ | 0.1379 | 0.3283 | 0.0437 | 2.8236 | cubic |
| 46 | $Ba_2TbCoO_6$ | 0.1213 | 0.2957 | 2.0995 | 2.3517 | cubic |
| 47 | $Ba_2DyCoO_6$ | 0.1195 | 0.2837 | 2.1404 | 2.4121 | cubic |
| 48 | $Ba_2HoCoO_6$ | 0.1192 | 0.2907 | 2.1736 | 2.4061 | cubic |
| 49 | $K_2YbTlF_6$ | 0.1326 | 0.4456 | 0.6244 | 0.5007 | cubic |
| 50 | $Cs_2KRbF_6$ | 0.1304 | 0.2202 | 1.1867 | -3.2444 | cubic |
| 51 | $Ba_2GdCoO_6$ | 0.1123 | 0.2608 | 2.2125 | 2.4466 | cubic |
| 52 | $Ba_3MnO_6$ | 0.1864 | 0.1119 | 1.6083 | 1.6057 | cubic |
| 53 | $Sr_2EuCoO_6$ | 0.1417 | 0.3675 | 2.3958 | 2.5145 | cubic |
| 54 | $Cs_2CoBr_7$ | 0.0988 | 0.1270 | 2.0530 | 0.0284 | cubic |
| 55 | $Cs_2YbTlCl_6$ | 0.0568 | 0.2595 | 0.7378 | 0.5575 | cubic |

| Number | Formula | EAH (eV/atom) | $E_{CD}$ (eV) | $E_{VD}$ (eV) | $E_F$ (eV) | Crystal system |
|---|---|---|---|---|---|---|
| 56 | $Ba_2CaVO_6$ | 0.1306 | 0.2642 | 1.6709 | 1.9929 | cubic |
| 57 | $Ba_2YCoO_6$ | 0.1977 | 0.2744 | 4.7175 | 1.3843 | cubic |
| 58 | $K_2GaPdF_6$ | 0.1432 | 0.1795 | 0.5051 | 1.3993 | cubic |
| 59 | $KCaSe_2$ | 0.2116 | 0.2305 | 2.3866 | 0.6279 | trigonal |
| 60 | $K_2RbSnF_6$ | 0.1533 | 0.3709 | 0.6343 | 1.6069 | cubic |
| 61 | $K_2MoBr_7$ | 0.1735 | 0.1344 | 0.6143 | -0.3767 | cubic |
| 62 | $EuBi_2Br_3$ | 0.2421 | 0.3915 | 3.2201 | -0.6868 | tetragonal |
| 63 | $Ca_2FeClO_4$ | 0.1713 | 0.3334 | 7.8364 | 3.7910 | tetragonal |
| 64 | $Ba_2SrVO_6$ | 0.1804 | 0.2257 | 1.6208 | 1.8001 | cubic |
| 65 | $NaCaSe_2$ | 0.1964 | 0.2119 | 2.7939 | 1.1561 | trigonal |
| 66 | $Cs_3YbF_6$ | 0.1379 | 0.0544 | 2.0857 | -2.1527 | cubic |
| 67 | $Cs_2KCaCl_6$ | 0.1206 | 0.0963 | 0.7126 | -1.6744 | cubic |
| 68 | $Ba_2YbCoO_6$ | 0.1637 | 0.3210 | 2.0420 | 2.0049 | cubic |
| 69 | $Cs_2FeCuCl_6$ | 0.0794 | 0.1294 | 1.2005 | 1.8650 | cubic |
| 70 | $Ba_2CaCoO_6$ | 0.1402 | 0.2674 | 2.0038 | 2.2325 | cubic |
| 71 | $Ba_2ErCoO_6$ | 0.1194 | 0.2867 | 2.1893 | 2.3695 | cubic |
| 72 | $Ba_2CaFeO_6$ | 0.1094 | 0.2687 | 1.8741 | 2.0555 | cubic |
| 73 | $K_2YbTlCl_6$ | 0.1177 | 0.2783 | 0.7429 | 0.3415 | cubic |
| 74 | $Rb_2EuTlCl_6$ | 0.0091 | 0.4079 | 0.7797 | -0.6957 | cubic |
| 75 | $K_2NaYbCl_6$ | 0.1334 | 0.2104 | 3.6718 | -0.5566 | cubic |
| 76 | $Ba_2ScCoO_6$ | 0.1329 | 0.2477 | 4.9532 | 1.7751 | cubic |
| 77 | $HfSbOs_1$ | 0.1045 | 0.1955 | 6.0432 | 6.8267 | cubic |
| 78 | $Rb_3YbCl_6$ | 0.1409 | 0.0809 | 1.5039 | -2.2511 | cubic |
| 79 | $Sr_3MnO_6$ | 0.2457 | 0.1863 | 1.6583 | 1.5877 | cubic |
| 80 | $Cs_2CuBr_6$ | 0.1054 | 0.4110 | 1.8586 | 0.1903 | cubic |
| 81 | $Sr_3FeO_6$ | 0.2208 | 0.2847 | 2.0682 | 1.6632 | cubic |
| 82 | $K_2NaYbF_6$ | 0.2102 | 0.0659 | 1.9239 | -3.3228 | cubic |
| 83 | $Ca_2W_2O_8F$ | 0.1346 | 0.4674 | 6.1011 | 0.2129 | monoclinic |
| 84 | $BaVO_4$ | 0.1906 | 0.3862 | 3.6871 | -0.0168 | triclinic |

| Number | Formula | EAH (eV/atom) | $E_{CD}$ (eV) | $E_{VD}$ (eV) | $E_F$ (eV) | Crystal system |
|---|---|---|---|---|---|---|
| 85 | $Cs_3(CeSe_3)_2$ | 0.0676 | 0.0883 | 2.5624 | 1.0686 | triclinic |
| 86 | $TiCr_2O_6$ | 0.1280 | 0.3981 | 0.6969 | 2.8777 | triclinic |
| 87 | $NaYbCl_4$ | 0.1511 | 0.3199 | 3.6308 | -0.2724 | triclinic |
| 88 | $Cs_7(PH)_2$ | 0.1106 | 0.1926 | 1.0177 | 1.7663 | triclinic |
| 89 | $K_4Mn(CuBr_4)_3$ | 0.0400 | 0.2902 | 4.1913 | 0.6806 | triclinic |
| 90 | $CsLaF_6$ | 0.1340 | 0.3044 | 2.6875 | -0.3419 | monoclinic |
| 91 | $Ce_2Mo_2O_{11}$ | 0.2226 | 0.4015 | 5.0650 | 1.2637 | triclinic |
| 92 | $Hg_5F_{13}$ | 0.0256 | 0.4220 | 6.0966 | -2.2052 | triclinic |
| 93 | $V_3SiClO_7$ | 0.0871 | 0.3219 | 7.8255 | 3.9843 | triclinic |
| 94 | $KGeCl_4$ | 0.1802 | 0.3861 | 0.7721 | 0.3748 | triclinic |
| 95 | $CsRe_2ICl_8$ | 0.1830 | 0.0708 | 1.2996 | 0.2212 | triclinic |
| 96 | $CsBaCu_2F_6$ | 0.0978 | 0.1380 | 1.4455 | 1.1712 | orthorhombic |
| 97 | $BaVCrO_6$ | 0.1173 | 0.4861 | 0.5228 | 0.8880 | triclinic |
| 98 | $Dy_3WO_8$ | 0.1441 | 0.2769 | 4.7536 | 0.9390 | orthorhombic |
| 99 | $K_2Eu(OF_3)_2$ | 0.2036 | 0.0932 | 0.0024 | -2.6264 | triclinic |
| 100 | $Na_3LiF_6$ | 0.0446 | 0.4276 | 0.7597 | -2.2045 | monoclinic |
| 101 | $Rb_2AgCl_6$ | 0.0807 | 0.4366 | 1.9581 | -0.3604 | monoclinic |
| 102 | $HgSeF_5$ | 0.1322 | 0.3246 | 0.7780 | -0.8981 | triclinic |
| 103 | $CsNa_2ZrI_8$ | 0.0973 | 0.2670 | 3.3709 | 0.8400 | triclinic |
| 104 | $CsTiF_3$ | 0.1999 | 0.3157 | 0.7482 | 5.0698 | monoclinic |
| 105 | $Y_3Se_5Br_4$ | 0.2074 | 0.2000 | 3.9765 | 0.0872 | triclinic |
| 106 | $Eu_2OF_4$ | 0.0968 | 0.4032 | 0.6030 | 1.0172 | triclinic |
| 107 | $BaSrF_6$ | 0.0886 | 0.2696 | 3.3238 | -1.5128 | tetragonal |
| 108 | $Cr_2As_3Se_8$ | 0.0201 | 0.4291 | 6.0308 | 3.3466 | monoclinic |
| 109 | $RbBaSe_2$ | 0.2195 | 0.2683 | 1.7284 | -0.0349 | trigonal |
| 110 | $Cs_3Te_4$ | 0.0973 | 0.2576 | 2.6671 | 1.7362 | triclinic |
| 111 | $K_3YbF_8$ | 0.2240 | 0.2560 | 2.4540 | -3.8948 | monoclinic |
| 112 | $KI_2$ | 0.0650 | 0.2589 | 5.3532 | 1.9796 | triclinic |
| 113 | $Na_3EuF_8$ | 0.0005 | 0.1460 | 2.7200 | -2.8007 | monoclinic |

| Number | Formula | EAH (eV/atom) | $E_{CD}$ (eV) | $E_{VD}$ (eV) | $E_F$ (eV) | Crystal system |
|---|---|---|---|---|---|---|
| 114 | $Cs_2RuBr_9$ | 0.0798 | 0.1878 | 3.7825 | -0.2755 | triclinic |
| 115 | $Cs_4Se_3$ | 0.2197 | 0.2238 | 1.0124 | 0.2784 | triclinic |
| 116 | $Rb_2CdAuF_6$ | 0.1181 | 0.3974 | 1.2345 | -0.2490 | cubic |
| 117 | $Cs_2SrF_6$ | 0.1624 | 0.2133 | 1.7071 | -3.0575 | triclinic |
| 118 | $Tl(MoO_4)_2$ | 0.1894 | 0.2443 | 6.2122 | 0.8070 | triclinic |
| 119 | $CsBaF_4$ | 0.0783 | 0.1413 | 1.7337 | -2.9746 | tetragonal |
| 120 | $Na_2CuRuF_6$ | 0.1966 | 0.0760 | 1.3196 | 1.8970 | monoclinic |
| 121 | $CsRhBr_6$ | 0.1314 | 0.3619 | 1.8514 | 0.1149 | triclinic |
| 122 | $KTm_2Sb_2Se_7$ | 0.0940 | 0.2075 | 4.6448 | 2.5352 | triclinic |

## Table S2. Complete list of validated n-type cold metal candidates

| Number | Formula | EAH (eV/atom) | $E_{CD}$ (eV) | $E_{VD}$ (eV) | $E_F$ (eV) | Crystal system |
|---|---|---|---|---|---|---|
| 1 | $K_3ReF_6$ | 0.2200 | 0.0000 | 0.1434 | 2.8012 | cubic |
| 2 | $U_2Cu_2Se_4$ | 0.1763 | 0.0000 | 0.4399 | 2.8206 | tetragonal |
| 3 | $Cs_2BaTbCl_6$ | 0.1334 | 0.0000 | 0.2473 | 3.9011 | cubic |
| 4 | $Cs_2YbGdCl_6$ | 0.0475 | 0.0000 | 0.3058 | 4.5077 | cubic |
| 5 | $Rb_3SeBr_5$ | 0.1630 | 1.1265 | 0.2024 | -1.2836 | tetragonal |
| 6 | $Cs_2YbTmCl_6$ | 0.1120 | 0.0000 | 0.2199 | 4.7484 | cubic |
| 7 | $Cs_2YbNbF_6$ | 0.2067 | 0.0000 | 0.2804 | 3.8216 | cubic |
| 8 | $Cs_2BaEuCl_6$ | 0.0159 | 0.0000 | 0.1156 | 2.7705 | cubic |
| 9 | $Rb_2TiCl_7$ | 0.1084 | 0.7024 | 0.3280 | -0.6678 | cubic |
| 10 | $Na_2PrF_7$ | 0.0530 | 1.9340 | 0.1356 | -3.2356 | hexagonal |
| 11 | $K_2SrEuF_6$ | 0.0532 | 0.0000 | 0.3525 | 2.8619 | cubic |
| 12 | $KTbCl_3$ | 0.2139 | 0.0000 | 0.3698 | 3.8273 | cubic |
| 13 | $Cs_2YbScF_6$ | 0.1369 | 0.0000 | 0.2573 | 5.4447 | cubic |
| 14 | $Cs_2BaGdCl_6$ | 0.0872 | 0.0000 | 0.3278 | 3.6428 | cubic |
| 15 | $Cs_2SrNbF_6$ | 0.2077 | 0.0000 | 0.2655 | 3.8547 | cubic |

| Number | Formula | EAH (eV/atom) | $E_{CD}$ (eV) | $E_{VD}$ (eV) | $E_F$ (eV) | Crystal system |
|---|---|---|---|---|---|---|
| 16 | $K_2SrTiF_6$ | 0.1614 | 4.2154 | 0.4743 | 3.8836 | cubic |
| 17 | $Na_3PrF_6$ | 0.1344 | 0.0000 | 0.2317 | -3.6052 | cubic |
| 18 | $EuFe_5O_8$ | 0.2384 | 0.0000 | 0.4011 | 3.6930 | trigonal |
| 19 | $Cs_2YbPrCl_6$ | 0.0855 | 0.0000 | 0.3281 | 4.1677 | cubic |
| 20 | $Rb_2NaTiCl_5$ | 0.1371 | 1.2278 | 0.3216 | 2.6226 | cubic |
| 21 | $Cs_3NaTe_2Cl_4$ | 0.1765 | 1.3062 | 0.2471 | 0.6754 | cubic |
| 22 | $Cs_2SrCeCl_6$ | 0.0864 | 0.0000 | 0.4568 | 4.5850 | cubic |
| 23 | $Cs_2MnH_7$ | 0.0521 | 1.2009 | 0.3002 | 1.0540 | cubic |
| 24 | $Cs_3NaCl_6$ | 0.0740 | 2.4334 | 0.2229 | -2.2028 | cubic |
| 25 | $Cs_2BaErCl_6$ | 0.1465 | 0.0000 | 0.2712 | 4.0040 | cubic |
| 26 | $K_2TlSnF_6$ | 0.1441 | 0.9156 | 0.4534 | 2.1135 | cubic |
| 27 | $Na_2TiCoF_6$ | 0.0999 | 0.0000 | 0.3708 | 3.6357 | cubic |
| 28 | $RbTbCl_3$ | 0.1838 | 0.0000 | 0.4108 | 4.0237 | cubic |
| 29 | $Rb_2YbErCl_6$ | 0.1253 | 0.0000 | 0.3066 | 4.3137 | cubic |
| 30 | $BaEuCl_3$ | 0.2205 | 0.0000 | 0.4698 | 3.9453 | hexagonal |
| 31 | $Cs_2BaEuCl_6$ | 0.0521 | 0.0000 | 0.0829 | 2.7726 | cubic |
| 32 | $Rb_3TiCl_5$ | 0.1500 | 1.0307 | 0.2754 | 2.4234 | cubic |
| 33 | $Cs_2BaLaCl_6$ | 0.1314 | 0.0000 | 0.4428 | 4.0717 | cubic |
| 34 | $Cs_3FeCl_6$ | 0.1196 | 0.6175 | 0.4705 | 0.5869 | cubic |
| 35 | $Cs_2KNaCoF_5$ | 0.1856 | 0.0000 | 0.1582 | 1.4014 | cubic |
| 36 | $K_2TiF_7$ | 0.1278 | 1.2033 | 0.3146 | -2.6696 | cubic |
| 37 | $Rb_3NaF_6$ | 0.0492 | 0.5279 | 0.2062 | -2.4913 | cubic |
| 38 | $Na_3MnCl_5$ | 0.1071 | 2.6135 | 0.2777 | -0.7632 | cubic |
| 39 | $Cs_2TiTlF_6$ | 0.1660 | 1.4276 | 0.1980 | 1.1415 | cubic |
| 40 | $Rb_2YbDyCl_6$ | 0.1190 | 0.0000 | 0.3359 | 4.2571 | cubic |
| 41 | $Cs_2BaPrCl_6$ | 0.1307 | 0.0000 | 0.2868 | 3.6784 | cubic |

| Number | Formula | EAH (eV/atom) | $E_{CD}$ (eV) | $E_{VD}$ (eV) | $E_F$ (eV) | Crystal system |
|---|---|---|---|---|---|---|
| 42 | $Cs_2TlF_7$ | 0.0546 | 1.3743 | 0.2393 | -2.6280 | cubic |
| 43 | $Rb_2RuAuCl_6$ | 0.1468 | 0.0000 | 0.3641 | 1.6882 | cubic |
| 44 | $Cs_2Nb_2Br_6$ | 0.1268 | 0.0000 | 0.4935 | 2.5848 | hexagonal |
| 45 | $K_2TlWCl_6$ | 0.1281 | 0.0000 | 0.4223 | 1.9383 | cubic |
| 46 | $K_2TiMnH_6$ | 0.2434 | 0.5231 | 0.3530 | 2.0811 | cubic |
| 47 | $Cs_2YbNbCl_6$ | 0.1478 | 0.0000 | 0.4802 | 3.1325 | cubic |
| 48 | $Cs_2SrGdCl_6$ | 0.0661 | 0.0000 | 0.4264 | 4.1789 | cubic |
| 49 | $Cs_2SrPrCl_6$ | 0.1034 | 0.0000 | 0.2307 | 3.9208 | cubic |
| 50 | $Cs_2YbHoCl_6$ | 0.1022 | 0.0000 | 0.2616 | 4.6766 | cubic |
| 51 | $Cs_2TlCu_2Cl_5$ | 0.0194 | 0.0000 | 0.2415 | 1.4093 | cubic |
| 52 | $SrNdFeCoO_6$ | 0.1729 | 0.8845 | 0.1535 | 3.2527 | cubic |
| 53 | $CeZr_3TlSe_8$ | 0.1024 | 0.0000 | 0.0616 | 3.6985 | triclinic |
| 54 | $Ti_2MnNbO_7$ | 0.0719 | 0.0000 | 0.4998 | 4.2190 | triclinic |
| 55 | $Sr_4Br_7$ | 0.0676 | 0.0000 | 0.4476 | 4.0495 | triclinic |
| 56 | $Cs_6IrAuCl_{10}$ | 0.0421 | 1.9832 | 0.2393 | 0.5206 | triclinic |
| 57 | $CsNa_2InF_6$ | 0.0138 | 6.4548 | 0.3826 | -2.8352 | trigonal |
| 58 | $CrP_2Cl_{13}$ | 0.0548 | 0.7735 | 0.0901 | 0.7892 | triclinic |
| 59 | $Rb_6SrCeCl_{12}$ | 0.1042 | 2.2581 | 0.1161 | -0.9921 | monoclinic |
| 60 | $Rb_2AlOF_6$ | 0.2461 | 0.6804 | 0.1185 | -1.9396 | triclinic |
| 61 | $Cu_2As_2Cl_7$ | 0.1537 | 0.0000 | 0.1238 | 1.1664 | triclinic |
| 62 | $TiI_2F_9$ | 0.1148 | 0.5458 | 0.1846 | -0.6562 | triclinic |
| 63 | $CsCe_3I_8$ | 0.0671 | 0.6716 | 0.1306 | 4.0857 | triclinic |
| 64 | $Na_2ZrPbF_{10}$ | 0.0239 | 3.3729 | 0.2557 | -3.1600 | triclinic |
| 65 | $Pr_4SiBr_7$ | 0.1206 | 0.0000 | 0.1175 | 1.7535 | triclinic |
| 66 | $Cs_2NbIF_6$ | 0.1360 | 0.8951 | 0.2108 | 0.2250 | monoclinic |
| 67 | $BaTi(ClF)_2$ | 0.1918 | 0.0000 | 0.1535 | 4.3505 | monoclinic |

| Number | Formula | EAH (eV/atom) | $E_{CD}$ (eV) | $E_{VD}$ (eV) | $E_F$ (eV) | Crystal system |
|---|---|---|---|---|---|---|
| 68 | CaCePSe | 0.1057 | 0.0000 | 0.3606 | 5.2432 | trigonal |
| 69 | $Rb_4VCl_7$ | 0.0391 | 0.0385 | 0.4846 | 0.2064 | monoclinic |
| 70 | $Cs_5Rb(MnCl_5)_2$ | 0.0108 | 2.7635 | 0.3603 | 0.1045 | triclinic |
| 71 | $Ce_5CCl_8$ | 0.0899 | 0.0000 | 0.4187 | 3.8776 | monoclinic |
| 72 | $Cs_5Nb_2F_{12}$ | 0.0725 | 0.5633 | 0.1226 | 0.8441 | triclinic |
| 73 | $Ti_2OF_5$ | 0.1842 | 3.3761 | 0.1640 | 1.3161 | triclinic |
| 74 | $ScTlCl_3$ | 0.2146 | 0.0000 | 0.4193 | 4.9824 | monoclinic |
| 75 | $Cs_6RbCr_2Br_{11}$ | 0.0818 | 1.3602 | 0.3195 | 0.4614 | triclinic |
| 76 | $Na_3GdF_6$ | 0.1868 | 0.0000 | 0.2688 | -1.8745 | monoclinic |
| 77 | $TiMn_2CdI_8$ | 0.0737 | 0.0000 | 0.4933 | 2.4928 | triclinic |
| 78 | $Cs_3C_2N_3$ | 0.0699 | 0.0000 | 0.4466 | 1.0025 | triclinic |
| 79 | $Na_2Mn_3VO_6$ | 0.0775 | 0.0000 | 0.1682 | 4.7790 | monoclinic |
| 80 | $Rb_2ZrHgF_6$ | 0.1287 | 0.0000 | 0.4126 | -0.9915 | cubic |
| 81 | $Ba_3CeBr_8$ | 0.0713 | 0.0000 | 0.4629 | 4.9212 | triclinic |
| 82 | $NbICl_8$ | 0.0593 | 2.0986 | 0.1926 | -0.8564 | triclinic |
| 83 | $Mo_3(AsCl_4)_2$ | 0.1938 | 0.0334 | 0.0685 | -0.1655 | triclinic |
| 84 | $Sc_2SBr_3$ | 0.2134 | 0.0000 | 0.3689 | 3.1644 | monoclinic |
| 85 | $Rb_2PtF_7$ | 0.1090 | 0.0445 | 0.3369 | -2.0210 | cubic |
| 86 | $Cs_3Er_2Cl_7$ | 0.1584 | 0.0000 | 0.4770 | 4.3760 | triclinic |
| 87 | $Sr_7Mn(SO)_4$ | 0.1793 | 0.0000 | 0.2225 | 2.0997 | triclinic |
| 88 | $Nb_2N_3OF_{12}$ | 0.0808 | 0.7279 | 0.3341 | -4.2838 | triclinic |

**Table S3. Complete list of validated np-type cold metal candidates**

| Number | Formula | EAH (eV/atom) | $E_{CD}$ (eV) | $E_{VD}$ (eV) | $E_F$ (eV) | Crystal system |
|---|---|---|---|---|---|---|
| 1 | $Cs_2UCuCl_6$ | 0.1908 | 0.0502 | 0.1112 | 4.5024 | cubic |
| 2 | $Cs_3KF_6$ | 0.1010 | 0.1875 | 0.1843 | -2.4476 | cubic |

| Number | Formula | EAH (eV/atom) | $E_{CD}$ (eV) | $E_{VD}$ (eV) | $E_F$ (eV) | Crystal system |
|---|---|---|---|---|---|---|
| 3 | $Cs_2TiCoF_6$ | 0.0678 | 0.1429 | 0.0830 | 3.2058 | cubic |
| 4 | $Tl_2Ni_2Cl_4$ | 0.1970 | 0.2084 | 0.3731 | 2.3081 | hexagonal |
| 5 | $K_3ReCl_6$ | 0.1755 | 0.2881 | 0.1314 | 2.1202 | cubic |
| 6 | $Cs_2TiCuCl_6$ | 0.1637 | 0.1322 | 0.0834 | 3.1331 | cubic |
| 7 | $K_2TiFeF_6$ | 0.0065 | 0.0589 | 0.1405 | 3.6614 | cubic |
| 8 | $Na_3EuF_6$ | 0.1051 | 0.1037 | 0.1015 | -1.8764 | cubic |
| 9 | $Rb_3TaCl_6$ | 0.1578 | 0.1244 | 0.2370 | 2.2875 | cubic |
| 10 | $Sr_2Pd_2F_6$ | 0.1551 | 0.2642 | 0.2286 | 2.0063 | cubic |
| 11 | $K_2PdF_7$ | 0.1133 | 0.0690 | 0.4489 | -2.5773 | cubic |
| 12 | $Cs_2YbTiCl_6$ | 0.0532 | 0.1391 | 0.2965 | 3.5518 | cubic |
| 13 | $Cs_2PdF_7$ | 0.0845 | 0.1391 | 0.2171 | -2.3962 | cubic |
| 14 | $K_2TiNiF_6$ | 0.0356 | 0.2166 | 0.3386 | 3.6806 | cubic |
| 15 | $Cs_2NaFeCl_6$ | 0.1658 | 0.3128 | 0.4179 | -1.0512 | cubic |
| 16 | $SrPr_2VO_6$ | 0.1277 | 0.2428 | 0.2256 | 4.1520 | cubic |
| 17 | $Rb_2PdF_7$ | 0.0968 | 0.2815 | 0.2783 | -2.7444 | cubic |
| 18 | $Rb_2TiCoF_6$ | 0.0154 | 0.1982 | 0.4443 | 3.8320 | cubic |
| 19 | $K_2RbEuCl_6$ | 0.1456 | 0.1927 | 0.0557 | -2.3383 | cubic |
| 20 | $K_2SrRhF_6$ | 0.2301 | 0.1137 | 0.1675 | 1.7712 | cubic |
| 21 | $K_2RbEuCl_6$ | 0.0353 | 0.0601 | 0.2239 | -2.3738 | cubic |
| 22 | $Rb_2TiCuCl_6$ | 0.1421 | 0.2216 | 0.1945 | 2.8617 | cubic |
| 23 | $Rb_3NbH_6$ | 0.1398 | 0.0880 | 0.3540 | 0.3711 | cubic |
| 24 | $Na_3OsF_6$ | 0.2283 | 0.1729 | 0.2589 | 1.6368 | cubic |
| 25 | $Cs_2CuRuF_6$ | 0.1587 | 0.0641 | 0.0750 | 2.1688 | cubic |
| 26 | $Cs_2TiTlF_6$ | 0.0985 | 0.2372 | 0.1678 | 3.9032 | cubic |
| 27 | $Cs_2Se_6$ | 0.1899 | 0.4995 | 0.2809 | -2.2358 | tetragonal |
| 28 | $BaIr_2F_{11}$ | 0.0005 | 0.0978 | 0.0628 | -0.2099 | triclinic |

| Number | Formula | EAH (eV/atom) | $E_{CD}$ (eV) | $E_{VD}$ (eV) | $E_F$ (eV) | Crystal system |
|---|---|---|---|---|---|---|
| 29 | $Ce_3Br_8$ | 0.0092 | 0.2752 | 0.1593 | 2.5882 | triclinic |
| 30 | $Hg_3PF_6$ | 0.1217 | 0.4351 | 0.4604 | -1.7733 | triclinic |
| 31 | $Cs_2OsIBr_5$ | 0.0208 | 0.2611 | 0.2058 | 0.9388 | monoclinic |
| 32 | $Na_2TiNbF_8$ | 0.1593 | 0.0866 | 0.0795 | 3.3847 | tetragonal |
| 33 | $Ho_7RuI_{12}$ | 0.0165 | 0.1061 | 0.4759 | 4.1065 | triclinic |
| 34 | $Rb_2YbAgF_6$ | 0.0501 | 0.2926 | 0.1981 | -1.0262 | monoclinic |
| 35 | $Eu_2O_2F_3$ | 0.1400 | 0.0833 | 0.4510 | 1.5764 | monoclinic |
| 36 | $K_6EuMn_3O_8$ | 0.0813 | 0.0536 | 0.0913 | 1.8412 | triclinic |
| 37 | $Ce_2TiNiO_6$ | 0.2144 | 0.0545 | 0.1563 | 7.6760 | trigonal |
| 38 | $K_5(CoCl_4)_3$ | 0.0539 | 0.1584 | 0.0726 | 0.5593 | triclinic |
| 39 | $KRb_3Ag_2F_{12}$ | 0.0112 | 0.0729 | 0.3107 | -2.9547 | triclinic |
| 40 | $Na_2TiNbCl_8$ | 0.0913 | 0.2401 | 0.3489 | 2.4505 | triclinic |
| 41 | $Cs_2KF_7$ | 0.0835 | 0.0678 | 0.0800 | -3.1409 | monoclinic |
| 42 | $Ce_4NCl_6O$ | 0.0730 | 0.0886 | 0.0734 | 6.6431 | triclinic |
| 43 | $HgIrF_8$ | 0.0541 | 0.4476 | 0.3944 | -2.5553 | triclinic |
| 44 | $Rb_2IrPdF_6$ | 0.2123 | 0.2831 | 0.3740 | 1.2458 | monoclinic |
| 45 | $Rb_2CaF_7$ | 0.2018 | 0.0874 | 0.2253 | -3.2065 | triclinic |
| 46 | $YbCuF_6$ | 0.1408 | 0.1323 | 0.1913 | -2.5946 | trigonal |
| 47 | $Cs_5CeF_{14}$ | 0.0375 | 0.0751 | 0.0896 | -2.6675 | triclinic |